\documentclass[11pt]{book}

\usepackage[dvips]{graphicx,color}
\usepackage{makeidx,observe,cosmology}

\makeauthorindex
\makeindex

\BookTitle{New Trends in Theoretical and Observational Cosmology}
\CopyRight{\copyright 2001 by Universal Academy Press, Inc.}

\begin{document}

\BookTitle{\itshape Proceedings of the 5th RESCEU International Symposium}
\CopyRight{US-02-01
}
\pagenumbering{arabic}

\chapter{
Brane World, Mass Hierarchy and the Cosmological Constant}

\author{%
Shoichi ICHINOSE\\
{\it 
Laboratory of Physics, 
School of Food and Nutritional Sciences, 
University of Shizuoka,
Yada 52-1, Shizuoka 422-8526, Japan
}}
%
%
\AuthorContents{S.\ Ichinose} 

\AuthorIndex{Ichinose}{S.}

\section*{Abstract}
The brane world based on the 6D gravitational model
is examined. 
It is regarded as a higher dimensional version of
the 5D model by Randall and Sundrum . 
The obtained analytic solution is checked by
the numerical method. 
The mass hierarchy is examined. 
Especially the {\it geometrical
see-saw} mass relation, between
the Planck mass,
 the cosmological constant, and the neutrino mass, is suggested. 
Comparison with the 5D model is made.

\section{Introduction}

The higher dimensional approach is a natural way
to analyze the 4D physics in the geometrical standpoint
The history traces back to the work by Kaluza and Klein. 
Stimulated by the recent
development of the string and D-brane theories, a new type
 compactification mechanism was invented by Randall and
Sundrum\cite{RS9905,RS9906}. 
The domain wall configuration in 5D space-time,
which is a kink solution in the extra dimension, 
is exploited.
The D-brane inspired model
has provided us with new possibilities for the extension
of the standard model, with or without the supersymmetry.
It has some advantages in 
the hierarchy problem and the chiral problem.
We present a 6D soliton solution, and show that it provides
a new dimensional reduction mechanism\cite{SI0012,SI0103}.

\section{Six Dimensional Model and Brane World Solution}

We consider the 6D gravitational theory 
with the 6D Higgs potential.
\begin{eqnarray}
S[G_{AB},\Phi]=\int d^6X\sqrt{-G} (-{\frac{1}{2}} M^4{\hat R}
- G^{AB}\partial_A\Phi^*\partial_B\Phi-V(\Phi^*,\Phi))
{\quad ,}
\label{model1}
\end{eqnarray}
where $
V(\Phi^*,\Phi)=\frac{\lambda}{4}(|\Phi|^2-{v_0}^2)^2+\Lambda,\ 
(X^A)\equiv (x^\mu,\rho,\varphi), \mu=0,1,2,3.$\ 
$x^\mu$'s are regarded as our world coordinates, whereas  
$(X^4,X^5)=(\rho,\varphi)$ the extra ones. 
$\rho$ and $\varphi$ are taken as in Fig.1.
\begin{figure}[t]
  \begin{center}
    \includegraphics[height=10pc]{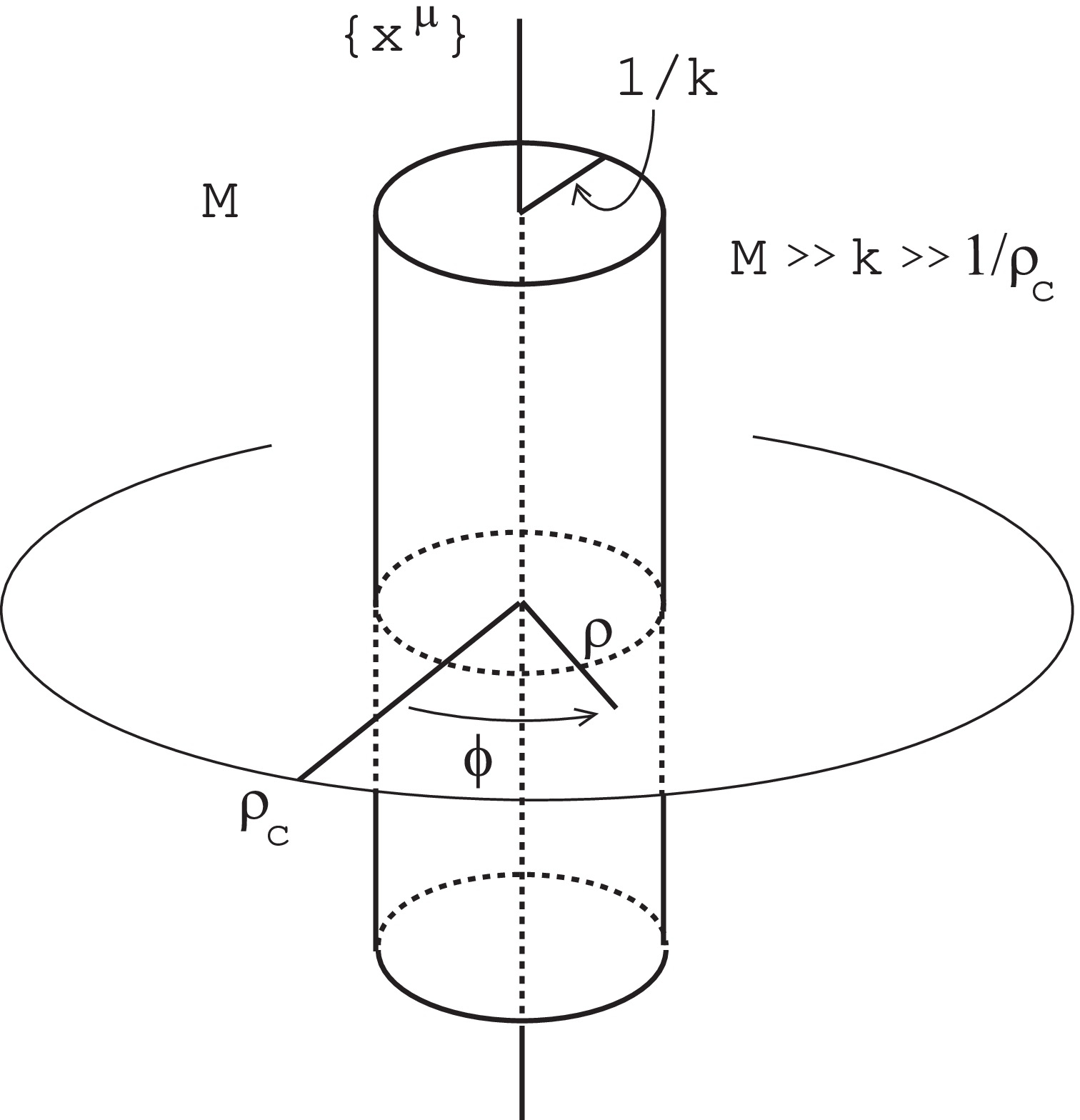}
    \includegraphics[height=10pc]{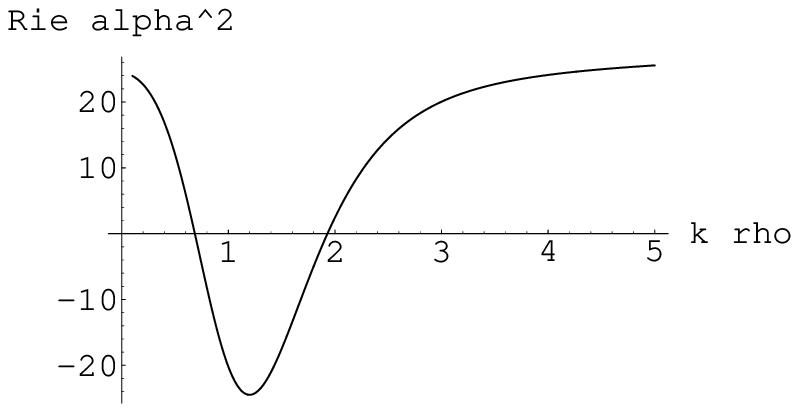}
  \end{center}
  \caption{[LEFT] The pole configuration.
  [RIGHT] (6D) Riemann scalar curvature ${\hat R}/\alpha^2$
in the 6th order approximation. 
The horizontal axis is $k\rho$. 
  }
\end{figure}
$M(>0)$ is the 6D Planck mass
and is regarded as the {\it fundamental scale} of this dimensional reduction
scenario. 
We take the line element:\ 
$
{ds}^2={\rm e}^{-2\sigma(\rho)}\eta_{\mu\nu} dx^\mu dx^\nu+{d\rho}^2
+\rho^2{\rm e}^{-2\omega(\rho)}d\varphi^2
$ 
($
0\leq \rho<\infty{\quad ,}\quad 0\leq\varphi<2\pi$),
where $\eta_{\mu\nu}=\mbox{diag}(-1,1,1,1)$. 
In this choice, the 4D Poincar{\' e}
invariance is preserved. Two "warp" factors ${\rm e}^{-2\sigma(\rho)}$ and
${\rm e}^{-2\omega(\rho)}$ appear.
The complex scalar field $\Phi$ is
periodic with repect to $\varphi$. 
Taking a simple case :\ 
$
\Phi_m(\rho,\varphi)= P(\rho){\rm e}^{im\varphi}
$, 
($
m=0,\pm1,\pm2,\cdots
$),\ 
the Einstein equation reduces to
$
3\sigma''-\frac{\sigma'}{\rho}+\sigma'\omega'+\omega''-(\omega')^2
+2\frac{\omega'}{\rho}
=2M^{-4}{P'}^2
$; 
$
-16{\sigma'}^2+4\frac{\sigma'}{\rho}-4\sigma'\omega'+4\sigma''
=2M^{-4}V
$; 
$4\sigma''-10{\sigma'}^2
=M^{-4}({P'}^2-m^2\frac{{\rm e}^{2\omega}}{\rho^2}P^2+V)
$. 

The boundary condition, 
at $\rho=\infty$(infrared region), for $P(\rho)$ is taken as :\ 
$
\rho\rightarrow\infty,\ P(\rho)\rightarrow +{v_0} .
$
As for $\sigma$ and $\omega$, 
we assume (from the "experience" in 5D Randall-Sundrum
model\cite{RS9905,RS9906,SI00apr,SI0107}) 
$\sigma'\rightarrow a(\mbox{const}), \omega'\rightarrow b(\mbox{const})$ as 
$\rho\rightarrow\infty$. 
Then, from the Einstein equations, we can deduce
$m=0$ and
$
\sigma'\rightarrow  \alpha\ , \omega'\rightarrow \alpha\ , 
\alpha=+\sqrt{\frac{-\Lambda}{10}}M^{-2}
$ as $\rho\rightarrow\infty$. 
Note that the present asymptotic requirement demands
the {\it isotropic property} around the $\rho=0$ axis
($m=0$), 
that is, the {\it pole} configuration. See Fig.1.
In the above result we must have the condition:
$
\Lambda<0
$
(Anti de Sitter). 
We can also fix the boundary condition  
at $\rho=+0$(ultra-violet region) based on the power-behavior assumption
and ${\hat R}$ regularity:\ 
As $\rho\rightarrow +0$,
the three functions $\sigma',\omega'$ and $P$ goes like
\ $\sigma'\rightarrow s\rho^a\ ,\ \omega'\rightarrow w\rho^b\ ,\ P\rightarrow x\rho^c$ where
$s,w$ and $x$ are some constants ($s\neq 0,w\neq 0$),\ and
$
a=b=1\ ,\ c=0\ ,\ s=\frac{\lambda}{16}(x^2-{v_0}^2)^2+\frac{\Lambda}{4}
$.

Let us take the following form for $\sigma'(\rho),\omega'(\rho)$ and $P(\rho)$
as a solution.
\begin{eqnarray}
\sigma'(\rho)=\alpha\sum_{n=0}^\infty\frac{c_{2n+1}}{(2n+1)!}\{\tanh (k\rho)\}^{2n+1}
\ ,\ \nonumber \\
\omega'(\rho)=\alpha\sum_{n=0}^\infty\frac{d_{2n+1}}{(2n+1)!}\{\tanh (k\rho)\}^{2n+1}
\ ,\ 
P(\rho)=v_0\sum_{n=0}^\infty\frac{e_{2n}}{(2n)!}\{\tanh (k\rho)\}^{2n}\ .
\label{sol9}
\end{eqnarray}
A {\it new mass scale} $k(>0)$ is introduced here and
 $1/k$ 
is the "thickness" of the pole. 
The parameter $k$, with $M$ and $\rho_c$(defined later), plays a central role
in this dimensional reduction scenario.
The distortion of 6D space-time by the
existence of the pole should be sufficiently small so that 
the quantum effect of 6D gravity
can be ignored and the present {\it classical} analysis is valid. This requires
the condition\cite{RS9905}:\ 
$
k\ll M
$. 

The infrared boundary conditions require
the coefficient-constants of (\ref{sol9})
to have the following constraints 
\begin{eqnarray}
1=\sum_{n=0}^\infty\frac{c_{2n+1}}{(2n+1)!}{\quad ,}\quad
1=\sum_{n=0}^\infty\frac{d_{2n+1}}{(2n+1)!}{\quad ,}\quad
1=\sum_{n=0}^\infty\frac{e_{2n}}{(2n)!}{\quad .}
\label{sol11}
\end{eqnarray}
We solve these constraints.
All coefficients 
are fixed except one {\it free} parameter $e_0$($=P(0)/v_0$).
The first two orders are concretely given as 
\begin{eqnarray}
\begin{array}{cc}
c_1=\frac{M^{-4}}{4k\alpha}\{ \frac{\lambda{v_0}^4(1-{e_0}^2)^2}{4}+\Lambda \},&
c_3=\frac{3\lambda^2{v_0}^6}{32 k^3\alpha}M^{-4}{e_0}^2(1-{e_0}^2)^2
+c_1(2+\frac{5\alpha}{k}c_1),
\\
d_1=-\frac{2c_1}{3}, e_0 : \mbox{free parameter},&
d_3=-\frac{4c_1}{3}(1+\frac{5\alpha c_1}{k}), e_2=-\frac{\lambda{v_0}^2}{4k^2}{e_0}(1-{e_0}^2).
\end{array}
\label{sol12}
\end{eqnarray}
The general terms $(c_{2n+1},d_{2n+1},e_{2n}), n\geq 2$ are
obtained in \cite{SI0012}. 
All coefficients are expressed
by four parameters $\lambda, {v_0}, \Lambda$ and $e_0$. 
The four ones have three constraints (\ref{sol11})
from the boundary condition at the infrared infinity. Hence
the present solution is {\it one-parameter family} solution. 
For an input value $e_0=-0.8$, the solution is concretely
obtained as in Fig.1[RIGHT] and in Fig.2[LEFT]. Furthermore
they are checked by the numerical method (Runge-Kutta) in Fig.2[RIGHT]
where no assumption is made about the form of the solution.
From the above solution (\ref{sol12}), we can easily estimate
the behavior of the vacuum parameters
($\Lambda,{v_0},\lambda$) near the 4D world limit (thin pole limit) as : 
$-\Lambda\sim M^4k^2$, 
${v_0}\sim M^{2}$, 
$\lambda\sim M^{-4}k^2$, 
as $k\rightarrow \infty$. 
As for the $k$-dependence, this result is the {\it same} as
the 5D model of Randall and Sundrum\cite{RS9905,SI00apr}.

\section{Physical constants and See-Saw relation}
Let us consider the case that
the 4D geometry is slightly fluctuating around 
the Minkowski (flat) space:\ 
$
{ds}^2={\rm e}^{-2\sigma(\rho)}g_{\mu\nu} dx^\mu dx^\nu+{d\rho}^2
+\rho^2{\rm e}^{-2\omega(\rho)}d\varphi^2,\ 
g_{\mu\nu}=\eta_{\mu\nu}+h_{\mu\nu} .
$
The gravitational part of 6D action (\ref{model1}) reduces to
4D action as
\begin{eqnarray}
\int d^6X\sqrt{-G}M^4{\hat R}\sim
M^4\int_0^{2\pi}d\varphi\int_{0}^{\rho_c}d\rho\,\rho{\rm e}^{-2\sigma(\rho)-\omega(\rho)}
\int d^4x\sqrt{-g}R+\cdots{\quad ,}
\label{para2}
\end{eqnarray}
where the {\it infrared regularization} parameter $\rho_c$
is introduced. $\rho_c$ specifies the {\it size} of the extra 2D space.
Using the asymptotic forms, $\sigma\sim \alpha \rho,\ \omega\sim \alpha \rho$ 
as $\rho\rightarrow \infty$ and
$\alpha=
\sqrt{\frac{-\Lambda}{10}}M^{-2}\sim k$ as $k\rightarrow\infty$, 
we can evaluate $M_{pl}$ as :\ 
$
{M_{pl}}^2\sim M^42\pi\int_{0}^{\rho_c}d\rho\,\rho\,{\rm e}^{-3\alpha\rho}
\sim \frac{M^4}{\alpha^2}\sim\frac{M^4}{k^2}
$,\ 
where we have used the {\it 4D reduction condition}:
$
k\rho_c\gg 1
$. 
(This result is {\it different} from
5D model\cite{RS9905,SI00apr}:\ 
${M_{pl}}^2\sim\frac{M^3}{k}$.)
Writing the above result as
$
M_{pl}/{M}\sim M/k
$,
we notice this mass relation is the {\it geometrical see-saw}
relation corresponding to the 2 by 2 matrix\ : 
($0,\ M$\ {\tt //}\ $M,\ M_{pl}$). 
This provides the geometrical approach to the see-saw mechanism
which is usually explained by the diagonalization of the (neutrino)
mass matrix. ( See a textbook\cite{MP91}.)
%
Similarly the 4D cosmological constant $\Lambda_{4d}$ is evaluated as
$
\Lambda_{4d}\sim 
\frac{\Lambda}{\alpha^2}\sim -M^4<0\ ,\ 
k\rho_c\gg 1\ .
$
Using the value $M_{pl}\sim 10^{19}$GeV , the "rescaled" cosmological
parameter ${\tilde \Lambda}_{4d}\equiv \Lambda_{4d}/{M_{pl}}^2$
has the relation:
$
\sqrt{-{\tilde \Lambda}_{4d}}\sim k\sim M^2\times {10}^{-19}\ \mbox{GeV}\ .
$
The unit of mass is GeV here and in the following. 
The observed value of $\sqrt{|{\tilde \Lambda}_{4d}|}$
is roughly ${10}^{-41}$.\cite{Perl99} 

Some typical cases are
1) ($k={10}^{-41}, M={10}^{-11}$),\ 
2) ($k={10}^{-13}, M={10}^3$)\ 
3) ($k=10, M={10}^{10}$)\ 
4) ($k={10}^4, M={10}^{11.5}$)\ 
and 
5) ($k={10}^{19}, M={10}^{19}$).
Cases 3) and 4) are moderate cases which are acceptable except
for the cosmological constant. 
At present any choice of ($k,M$)
looks to have some trouble if we take into account the cosmological
constant. We consider the observed cosmological constant value
should be explained by some unknown mechanism. 
As in the Callan and Harvey's paper\cite{CH85}, we can
have the 4D {\it massless chiral} fermion bound to the wall
by introducing 6D {\it Dirac} fermion $\psi$ into (\ref{model1}).
If we {\it regulate} the extra axis by the finite range $0\leq y\leq \rho_c$,
the 4D fermion is expected to have a small mass 
$m_f\sim k{\rm e}^{-k\rho_c}$. 
If we take case 4) 
and regard the 4D fermion as a neutrino ($m_\nu\sim {10}^{-11}-{10}^{-9}\mbox{GeV}$),
we obtain
$\rho_c=(3.45-2.99)\times {10}^{-3}\mbox{GeV}^{-1}$. 
When the quarks or other leptons ($m_q,m_l\sim 10^{-3}-10^2\mbox{GeV}$) 
are taken as the 4D fermion, 
we obtain $\rho_c=(1.61-0.461)\times {10}^{-3}$GeV$^{-1}$.
It is quite a fascinating idea
to {\it identify the chiral fermion zero mode bound to the pole
with the neutrinos, quarks or other leptons}.

\begin{figure}[t]
  \begin{center}
    \includegraphics[height=7pc]{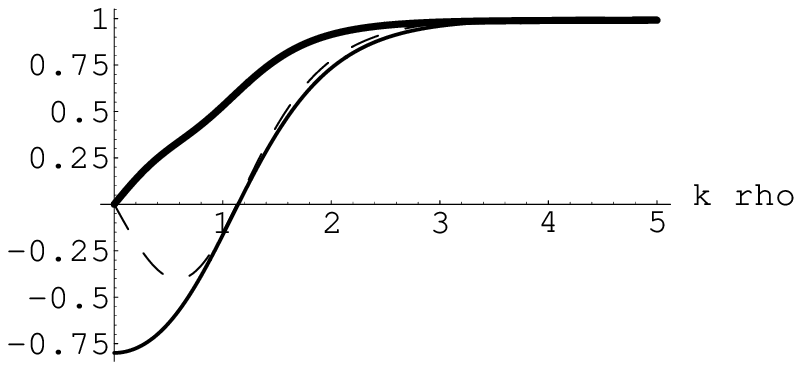}
    \includegraphics[height=7pc]{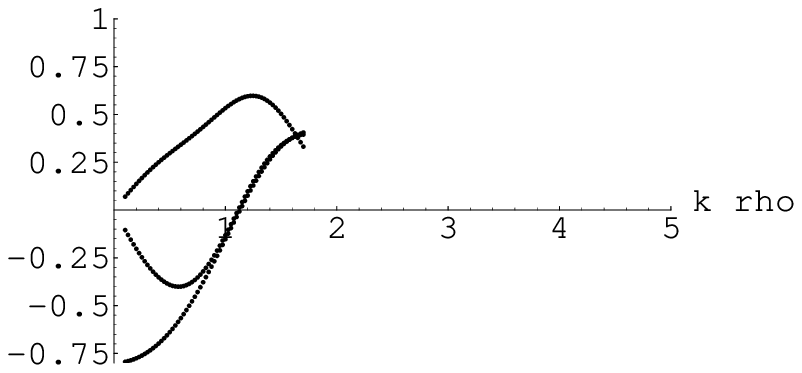}
  \end{center}
  \caption{Horizontal axis: $k\rho$. 
  [LEFT] The analytic results of $\omega'/\alpha$(bold line), 
  $\sigma'/\alpha$(dashed line) and $P/{v_0}$(normal line). $e_0=-0.8$. 
The graphs are depicted in the 6-th order approximation.
[RIGHT] The numerical results for $\omega'/\alpha$(top), $\sigma'/\alpha$(middle) and
$P/{v_0}$(down). They are obtained by Runge-Kutta method.
The initial point is $k\rho=0.1$. The initial values are borrowed
from the analytical results.
  }
\end{figure}

\section{Discussion and conclusion}
We add some numerical fact about
the see-saw relation.
In the case 1), the value of $M$ 
($\sim 10^{-11}$GeV) is the order of the
neutrino mass. 
This choice looks ridiculous because the space-time
behaves as six dimensional at the cosmological scale. 
The choice is, however, attractive in that 
it gives the right value of the cosmological constant. 
If this numerical fact is not accidental and has meaning,
it says the cosmological size is related to the neutrino mass
when it is "see-sawed" with the Planck mass.
These three fundamental scales could be geometrically related.
We hope the results of the present hierarchy model will lead to
develop further rich possibilities in the cosmology.




\begin{thebibliography}{99}
\bibitem{RS9905} 
L.Randall and R.Sundrum, {
Phys.Rev.Lett. {\bf 83}(1999)3370,hep-ph/9905221
}
\bibitem{RS9906} 
L.Randall and R.Sundrum, {
Phys.Rev.Lett. {\bf 83}(1999)4690,hep-th/9906064
}
\bibitem{SI0012}   
S.Ichinose,{Univ.of Shizuoka preprint,
US-00-11, hep-th/0012255,
"Pole Solution in Six Dimensions and Mass Hierarchy"}
\bibitem{SI0103} 
S. Ichinose, {Phys.Rev. {\bf D64}(2001)025012
[Erratum ibid.{\bf D65}(2001)029901], hep-th/0103211}
\bibitem{SI00apr}   
S.Ichinose,{Class.Quant.Grav.{\bf 18}(2001)421,hep-th/0003275 
}
\bibitem{SI0107}   
S.Ichinose,{Class.Quant.Grav.{\bf 18}(2001)5239,hep-th/0107254
}
\bibitem{MP91} 
R.N.Mohapatra and P.B.Pal,{"Massive Neutrinos in Physics and Astrophysics",
World Scientific, Singapore,1991}
\bibitem{Perl99} 
Perlmutter et al.,Astro.Phys.Journ.{\bf 517}(1999)567.
\bibitem{CH85} 
C.G.Callan and J.A.Harvey,{Nucl.Phys.{\bf B250}(1985)427}
\end{thebibliography}
\end{document}